\documentclass[]{aa}
\usepackage{latexsym,graphicx,natbib,amssymb,array,mathtools,ulem,txfonts,listings}
\newcommand{\bipos}{\textsc{BiPoS1}}
\newcommand{\ocen}{$\omega$~Cen}

\newcommand{\msun}{M_{\odot}}
\newcommand{\myr}{\msun\;{\rm yr}^{-1}}
\newcommand{\mpc}{M_{\odot}\;{\rm pc}^{-3}}

\newcommand{\fb}{f_{\rm b}}
\newcommand{\nb}{N_{\rm b}}
\newcommand{\ns}{N_{\rm s}}
\newcommand{\ncms}{N_{\rm cms}}
\newcommand{\rhoh}{\rho_{\rm h}}
\newcommand{\mbar}{\bar{m}}
\newcommand{\mecl}{M_{\rm ecl}}
\newcommand{\meclmin}{M_{\rm ecl,min}}
\newcommand{\meclmax}{M_{\rm ecl,max}}
\newcommand{\lP}{\log_{10}\rm P}
\newcommand{\Ekin}{E_{\rm kin}}

\newcommand{\FeH}{[Fe/H]}

\newcommand{\kms}{{\rm~km~s}^{-1}}

\newcommand{\tcr}{t_{\rm cr}}

\newcommand{\rh}{r_h}

\begin{document}

\title{A possible solution to the Milky Way's binary-deficient retrograde stellar population}
\subtitle{Evidence that $\omega$~Centauri has formed in an extreme starburst}
\titlerunning{$\omega$ Cen and binaries on retrograde orbits}

\author
{
  Michael Marks\inst{1} \and Pavel Kroupa\inst{2,3} \and Jörg Dabringhausen\inst{3}\\
}       

\institute{Erzbischöfliches Clara-Fey-Gymnasium Bonn-Bad Godesberg, Rheinallee 5, 53175 Bonn, Germany\\\email{astro.michi@yahoo.com} \and
           Helmholtz-Institut f\"ur Strahlen- und Kernphysik, University of Bonn, Nussallee 14-16, 53115 Bonn, Germany\\\email{pavel@astro.uni-bonn.de} \and Charles University in Prague, Faculty of Mathematics and Physics, Astronomical Institute, V  Hole\v{s}ovi\v{c}k\'ach 2, CZ-18000 Praha, Czech Republic\\\email{joerg@sirrah.troja.mff.cuni.cz}}

\date{Received <date>; accepted <date>}

\abstract{The fraction of field binaries on retrograde orbits about the Milky Way is significantly lower compared to its prograde counterpart. Chemical and dynamical evidence suggests that the retrograde stellar population originates from $\omega$~Centauri, which is either the most massive globular cluster (GC) of the Milky Way or the putative core of a former dwarf galaxy.}
{Star formation conditions required to produce the retrograde binary population are constrained assuming that the retrograde stellar population originates from $\omega$~Centauri's progenitor.}
{We match the observed low binary fraction with dynamical population synthesis models, including a universal initial binary population and dynamical processing in star clusters, making use of the publicly available binary population synthesis tool \bipos.}
{It is found that either the GC progenitor of \ocen\  must have formed with a stellar density of $\approx10^8\mpc$ or the $\omega$~Centauri dwarf galaxy's progenitor star cluster population must have formed in an extreme starburst with a star formation rate exceeding $1000\myr$ and probably a top-heavy embedded-cluster mass function with suppressed low-mass cluster formation. The separation and mass-ratio distribution for retrograde field binaries are predicted for comparison with future observations.}
{A viable solution for the deficiency of binaries on retrograde orbits is presented, and star formation conditions for $\omega$~Centauri as well as orbital parameter distributions for the Milky Way's retrograde binary population are predicted. The dwarf galaxy origin for $\omega$~Centauri is tentatively preferred within the present context.}


\maketitle

\section{Introduction}
\label{sec:retrograde}
\citet{c05} investigated the properties of spectroscopically selected field binaries observed in the Carney-Latham survey \citep[see references in][]{c05} for binary periods $P=1.7$~to~$7500$~days (corresponding to a semi-major axis range $a=0.03$~to~$8$~AU for a system mass of $1.2\msun$) and masses of the primary component $m_1\approx0.4$~to~$1\msun$. Amongst their sample of $1406$~systems, they find a binary fraction of $\fb=21.4\pm1$~per~cent, which compares excellently to the $\fb\approx21\pm3.5$~per~cent identified by \citet{DuqMay91} for G-dwarf field binaries over the same period range (see Fig.~\ref{fig:pdists}). About half of their systems with metallicities \FeH$<-1$ have retrograde orbits with disc velocities $V<-220\kms$ (the Sun has $V=0\kms$ in this co-rotating coordinate system).\footnote{The fraction of systems moving on retrograde orbits for higher metallicities is of the order  of $10$~per~cent only, which is not surprising as the metal-rich stars in their sample will be mainly associated with the disc of the Milky Way.} Interestingly, while the fraction for the $V>-220\kms$ prograde and metal-poor subsample has $\fb=28\pm3$~per~cent, they find the binary fraction among the retrograde, metal-poor systems to be only $\fb=10\pm2$~per~cent.

\citet{c05} attempted to determine where this deficiency of binaries on retrograde orbits stems from. They discuss various mechanisms that could possibly be responsible for this finding -- such as tidal disruption by massive objects in the Milky Way and the evaporation of single stars from clusters into, or from, clusters on retrograde orbits -- but rule out most of them based on qualitative and simple quantitative arguments. An attractive notion that remains is that the retrograde population of stars might be connected to $\omega$~Centauri, which would either be the most massive known globular cluster (GC) in the Milky Way \citep[$M=4\times10^6\msun$,][]{dr2013} or the hypothetical core of a former dwarf galaxy that was accreted by the Milky Way,   since \ocen\  circulates the Milky Way on a retrograde orbit as well. \citet{c05} put forward tentative chemical evidence of that connection: The abundance of the s-process elements barium ([Ba/Fe]) and yttrium ([Y/Fe]) is enhanced in \ocen\  stars compared to other Galactic GCs in the metallicity range \FeH$=-1.0$~to~$-1.6$. For stars and binaries on strongly retrograde orbits ($V<-300\kms$), the authors find evidence for a bimodal distribution in yttrium abundance \citep[Fig.~18 in][]{c05}\footnote{A cautionary note regarding the low-number statistics here: This finding is based on seven stars and one binary from their sample that fulfill the selection criteria ($V<300\kms$ and \FeH$\in[-1.6,-1.0]$); a Kolmogorov-Smirnov test performed by \citet{c05} did not reveal any significant difference between the prograde and retrograde sample.}. The separation between the two groups of systems is found to be $0.35\pm0.06$~dex, the very same magnitude that distinguishes giant stars in \ocen\  from those in other Galactic GCs. \citet{c05} also argue for a difference in barium abundance, although it is less pronounced and the observed magnitude of the difference ($0.2$~dex) is only half of what would be expected from the comparison of $\omega$ Cen with other GCs.

In addition to its unclear nature, \ocen\  is also a peculiar object in terms of its binary content. It contains virtually no binaries. While most Milky Way GCs share sub-field binary fractions \citep{m2012acs}, \ocen\  is an extreme among them. \citet{e1995wcen} estimate an upper limit to the equal-mass binary fraction of only $\approx5$~per~cent in the mass range $0.4-0.7\msun$ at a distance of $3-5$~half-mass radii from the core. \citet{mayor1996wcen} surveyed $310$ giant stars and estimate the global binary fraction to be as low as $3-4$~per~cent, assuming a similar period distribution as for nearby G dwarfs in the field. As the binary fraction is mostly seen decreasing from the core to the halo of star clusters \citep{ml1986m67,r1998pleiades,m2012acs}, the binary fraction in the core of \ocen\  might be somewhat larger. However, the opposite of this trend ($\fb$ rising in the cluster halo) has been observed for F-type stars in the young ($15-30$~Myr) Large Magellanic Cloud star cluster NGC~1818 \citep{deg2013ngc1818,li2013ngc180518}. \citet{g2013} show this to be a possible intermediate state of star cluster evolution originating from cluster disruption in a tidal field and dynamical mass segregation at a relatively young dynamical age of a cluster (about after one half-mass median two-body relaxation time), but they also find this trend to be wiped out after $\approx4-6$ median two-body relaxation times. If \ocen\  took this course of evolution, the common decreasing trend of $\fb$ with increasing radius is expected to be restored at its present age of 11.5~Gyr \citep{fb2010}.

\citet{g2013} further show that at some point before the usual trend is re-established, a bimodal distribution in terms of the binary fraction is reached. Binaries that were dynamically processed near the cluster centre move outwards when the core is re-filled with the, on average, more massive binaries farther out due to mass segregation. This process allows binaries to be found in the cluster halo and in the core, with a binary-deficient gap in between. The outer binaries will eventually be tidally stripped from the cluster and become part of the field, while those that stay in the cluster will be further processed.

This is reminiscent of findings of a bimodal distribution of blue straggler stars \citep{mapelli2006,lanzoni2007,kls2009,ferraro2012,beccari2013,sf2019}, thought to be objects descended from binary evolution. Their numbers drop up to a few core radii while rising again towards the cluster halo in objects such as M3, M5, M55 (NGC~6809), 47Tuc, and NGC~6752. Once more, \ocen\  is  deviant from those clusters in that it shows a rather constant blue straggler fraction throughout its extent \citep{mapelli2006}. \citet{hg2017} suggest, however, that the bimodal distribution is a transient feature and independent evidence that \ocen\  might have evolved beyond such a state.

Thus, the apparent absence of binaries inside \ocen\  and the low binary fraction that is nonetheless larger on retrograde orbits in the field than inside \ocen\  may be qualitatively consistent with the perception that binaries on retrograde orbits might indeed be associated with \ocen. It can also be interpreted as another indicator, in addition  to the chemical evidence, for their possible connection.

Lastly, \cite{im2004}  also argue that \ocen\  can plausibly be identified with a stripped dwarf elliptical galaxy. They succeeded in reproducing the observed surface density profile by using N-body computations of a tidal stripping scenario, lending further support to this scenario.

The present contribution seeks to explain the binary deficiency on retrograde orbits if \ocen\  or its former host galaxy was indeed the source of this deficiency. This will allow its possible star formation conditions to be uncovered. To this end, the publicly available Binary Population Synthesiser\footnote{Downloadable from \texttt{https://github.com/JDabringhausen/BiPoS1}} \citep*[\bipos;][and see the appendix]{dmk2021} is utilised.

\section{Modelling the retrograde binary population}
\subsection{A universal initial binary population}
\label{intro:binaries}
\begin{figure}
 \includegraphics[width=0.48\textwidth]{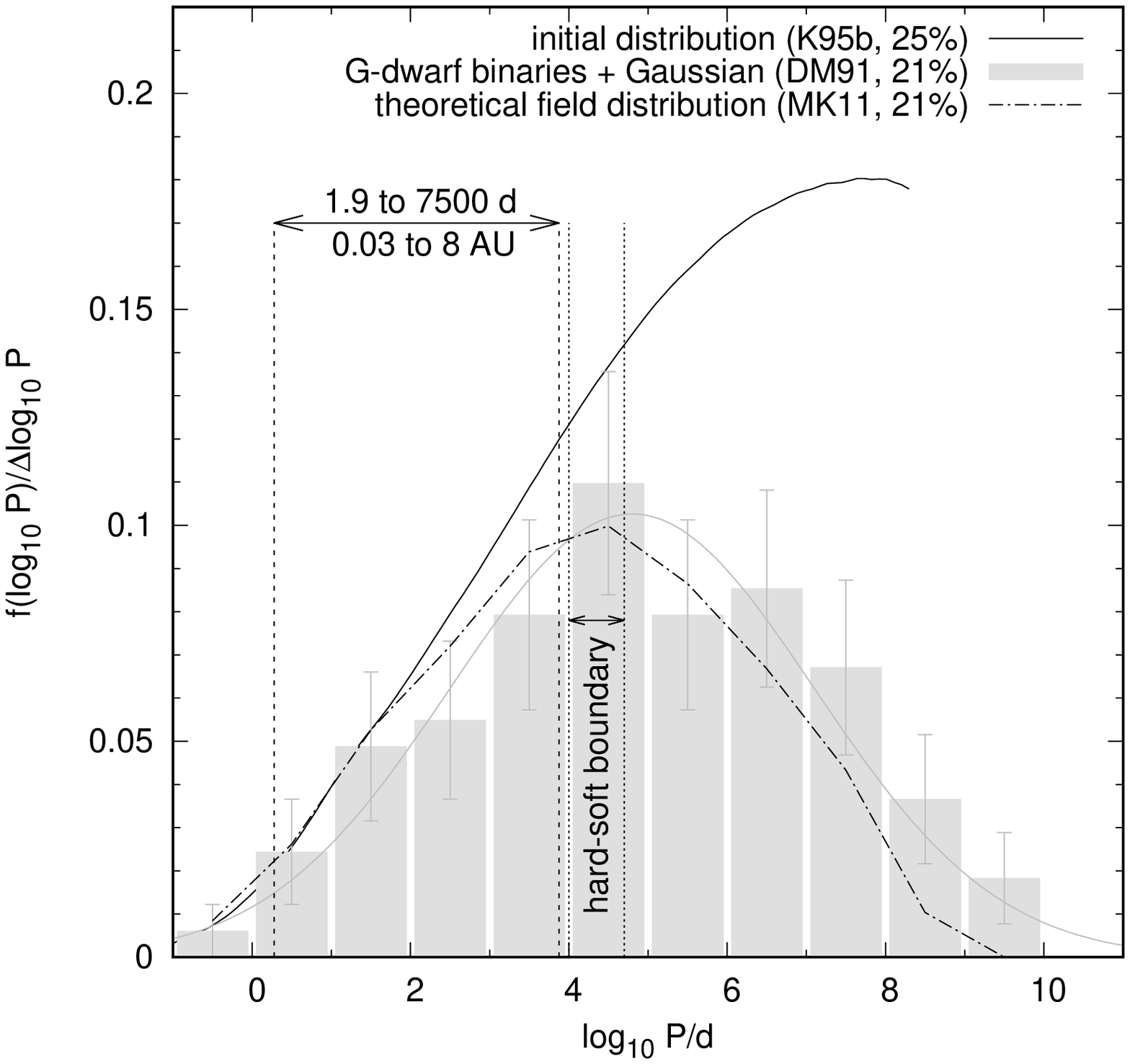}
 \caption[Binary period distributions]{Initial period distribution in star clusters \citep[solid line;][]{k95c,k95b}, the observed field G dwarf \citep[grey histogram plus Gaussian distribution;][]{DuqMay91}, and a theoretical binary period distribution for Galactic G dwarfs \citep[dash-dotted line;][]{mk11}. The period range between the dashed lines is the range of binary periods accessible in the spectroscopic survey by \citet{c05}. The percentages in the legend show the binary fraction within this range for the respective distribution. The region between the dotted lines is the location of the hard-soft boundary for a typical present-day GC ($M=10^5\msun$, $\rh=3$~pc; see Sect.~\ref{sec:hardsoft}). We note that $f(log_{10} P)=N_b(log_{10} P)/N_{cms}$, where $N_b(log_{10} P)$ denotes the number of binaries in the orbital period bin $log_{10} P$.}
 \label{fig:pdists}
\end{figure}
Stars generally do not form in isolation but in clusters of stars originating from giant molecular clouds \citep{ll2003,b2010,freeman2017,krause2020} where nearly every freshly hatched star carries a companion. Observations show that stars form predominantly as binary systems in sub-parsec density maxima inside those molecular clouds \citep{k95a,k95c,k95b,porras2003,mk11,megeath2016}, the properties of which change with galactocentric distance as a result of the varying local gas density \citep{pflamm2008,pflamm2013,miville2017,djordjevic2019}.

Therefore, binaries are common in all kinds of astrophysical objects \citep{dk2013rev}. Binary fractions are defined as
\begin{equation}
 \fb=\frac{\nb}{\ncms}\;,
 \label{eq:fb}
\end{equation}
where $\nb$ and $\ncms=\nb+\ns$ denote the number of binaries and the sum of all single and binary stars in a sample, respectively. A `centre-of-mass system' (cms) here refers to either a single star or a binary. Higher-order hierarchical multiples are much less frequent and are not considered in this contribution for the following two reasons. 

First, if star formation produced a significant fraction of higher-order multiples, then these would decay within a system-internal crossing time ($10^4-10^5\,$yr), typically leaving a binary and a number of single stars (particularly so in clustered regions). Thus, triple systems would decay to a binary and a single star, while quadruples would decay to a binary and two single stars, leaving, within $10^4-10^5\,$yr,  binary fractions of 0.5 and 0.33, respectively. Since dynamically unevolved star-forming regions (such as Taurus-Auriga) are observed to have binary fractions near 100~per cent at an age of about $1\,$Myr, this implies that the star formation process to a large extent produces binary systems and to some extent hierarchical higher-order systems that do not decay within 1 Myr \citep{gk2005}.

Second, of the observed higher-order multiple systems, some, and perhaps most, are likely to be the remnants of dissolved star clusters \citep{fuentemarcos1998}, reducing the fraction of stars formed in hierarchical multiple systems.

 Binary fractions are observed to lie in the range $10-50$~per~cent in GCs \citep{s2007,m2012acs} but may be as high as $70$~per~cent in open clusters \citep{s2010}. A little fewer than half of all systems in the Milky Way's Galactic field are binaries \citep{DuqMay91,r2010}.

It has been argued that these variations among different present-day binary populations are due to variations in their respective environment-dependent initial binary population (IBP), which is a trivial assertion \citep{King2012b,pm2014}. That the binary fraction for short-period ($P\lesssim7500$~d) binaries in the field has been demonstrated to be independent of metallicity \citep{c05}, suggesting little to no variation of binary properties over cosmic time, disfavours such a scenario, at least for said short-period binaries. Alternatively, the observed differences originate from a universal IBP \citep[the binary universality hypothesis;][]{k2011,kp2011} and subsequent environment-dependent dynamical evolution through interactions between stellar systems. The importance of dynamical modification of binary populations in star clusters has been reported in many contributions \citep{heggie1984,hut1985,heggie1988,mhm1990,mhm1991,heggie1991,hmv1991,ha1992,hmr1992,hut1992,heggie1994,mh1994,k95a,k95c,k95b,gs2000,kah2001,pz2001,f2003,i2005,thh2007,s2008,fir2009,pgkk2009,kaczmarek2011,mko11,pg2012,leigh2015,hong2015,lucatello2015,hong2016,wv2017,belloni2017}. Gravitational encounters dissolve the widest (`soft') binaries easily, such that an initially binary-dominated population is quickly diminished within a few crossing timescales \citep[of the order of a few megayears or less;][]{mko11} in a young and dense star cluster, which is termed `stimulated evolution'.

A universal IBP \citep[see Fig.~\ref{fig:pdists}]{k95c,k95b} matches observed pre-main-sequence and Class I stellar multiplicity data \citep{kp2011}. It has been demonstrated that subjecting this initial population to stellar-density-dependent stimulated evolution successfully reproduces not only the binary frequency but also binary distributions observed in most\footnote{Except USco-B in the Scorpius Centaurus OB association, whose observed period distribution is rising at long periods, i.e. where the soft binaries are, and is thus not compatible with any model of stimulated evolution \citep[as discussed in][]{mk12}.} young star formation regions, which have been thoroughly surveyed for their binary properties \citep{mk12,m14}. \citet{King2012b} have argued that this model is an incorrect description of observational reality, but \citet{m14} show that their conclusion would only hold up if star clusters and young star formation regions had not been significantly denser in the past; as such, a universal IBP plus dynamical processing of a denser progenitor remains a valid model.

Additionally, the Galactic field population for late-type stars can be reproduced with the same ansatz. In the \citet{mk11} models it is assumed that all stars currently in the field originated from dissolved embedded star clusters after allowing for some time of dynamical processing of the IBP inside each birth cluster \citep{k95a,k95c,k95b,g2010,liu2019}. Weighting each stellar population with a power-law embedded cluster mass function (ECMF; see Eq.~\ref{eq:xiecmf}), the remaining single and binary star populations from all clustered environments are tallied up to form a galaxy's field population. Once again, this model simultaneously matches both observed binary fractions and orbital parameter distributions (OPDs; period, mass ratio, and eccentricity) for field binaries of different spectral types in the Milky Way with a single set of model parameters.

Hence, there exists a universal IBP with a formal multiplicity fraction of unity that reproduces observations and is able to make predictions for different stellar populations both in terms of binary fractions and their OPDs. Additionally, the present-day binary fraction correlates with cluster age and luminosity \citep[i.e. mass;][]{s2007,m2008,s2010,m2012acs}, which could possibly not be expected if the IBP varied strongly from environment to environment. Thus, there does not seem to be a strong case for variations in the IBP.

The \citet{mko11} and \citet{mk11} methods used in this work have been collated in the command-line tool \bipos~described  in \citet*[and see the Appendix]{dmk2021}.

\subsection{Producing very low binary fractions dynamically}
\label{sec:hardsoft}
In order to find a very low binary fraction in any stellar system, many more single stars than binaries need to be present. Given the universal IBP (Fig.~\ref{fig:pdists}), this requires either that stimulated evolution is efficient in \ocen's star cluster progenitor or that in the \ocen\  galaxy's star cluster population those star clusters in which stimulated evolution was efficient were the dominant contributors to the field population.

A particular difficulty here in reproducing $\fb=10$~per~cent in the observed period range through stimulated evolution in star clusters is that \citet{c05}'s spectroscopically observed binaries lie on the `hard' side of the so-called hard-soft boundary. The binding energy of such binaries exceeds the kinetic energy of a typical cluster member, $E_b\gg\Ekin = \frac{1}{2}\mbar\sigma^2$, where $\mbar\approx0.4\msun$ is the average mass in the considered stellar population with the canonical stellar initial mass function \citep[IMF;][]{k2001} and $\sigma$ is its velocity dispersion. These binaries are difficult to dissolve through stimulated evolution as their cross-section is smaller than for soft binaries ($E_b\ll\Ekin$) and they tend to further harden during encounters \citep[Hills-Heggie law;][]{hills1975,heggie1975}. In the intermediate region around the hard-soft boundary ($E_b\approx\Ekin$), binary disruption is a particular stochastic process \citep{k95a,pg2012}. This is because the above definition of hard and soft binaries dilutes since it relies on average values. For example, a binary classified as hard may actually be soft in an energetic encounter. For a typical (present-day) GC of mass $M=10^5\msun$ and $\rh=3$~pc, the location of the hard-soft boundary is estimated as follows. For these values, the velocity dispersion in a star cluster following a Plummer density profile is
\begin{equation}
\sigma=\sqrt{s^2\frac{GM}{2\rh}}\approx5\frac{\rm pc}{\rm Myr}\;,\nonumber
\end{equation}
with $s=0.88$ being a structure factor and $G$ the gravitational constant. Therefore, the hard-soft boundary lies at a binary-internal binding energy of $E_b=5.8\msun\;{\rm pc}^2\;{\rm Myr}^{-2}$. For a binary system-mass $M_{\rm cms}=1.2\msun$, this corresponds to $\lP(d)=4-4.7$, depending on the mass ratio, $q$, of the components ($q=0.1$ or $1$ was used to estimate the lower and upper limit; see Fig.~\ref{fig:pdists}). \citet{c05} observed periods in the range $\lP(d)\approx0.3-3.9$.

Thus, while the binary fraction is rather simple to reduce through stimulated evolution for soft binaries, the binary fraction among hard binaries does not shrink at the same rate. It is, however, cautioned that the $\fb$ among hard binaries is lowered even if no hard binary is broken up. The hard binary fraction decreases by dissolving soft binaries since each disrupted binary increases $\ncms$ in Eq.~(\ref{eq:fb}) by one. This consideration alone, however, is not sufficient to reduce the binary fraction to the desired number.

A more important notion is that the so-estimated location of the hard-soft boundary refers to present-day GCs. As the density of any observed cluster has likely been larger in its past (Sect.~\ref{sec:singrequire}), the hard-soft boundary might have been located at lower values for $\lP$.

Under which conditions a low binary fraction as observed might have nevertheless been achieved is discussed in the following subsections.

\subsection{\ocen\  as a single cluster}
\label{sec:singrequire}
\citet{mko11} show the binary fraction in initially binary-dominated star clusters to evolve in the first few megayears on the initial crossing timescale, $\tcr\propto\rhoh^{-1/2}$, until the binary population finds an equilibrium with the cluster population. It is thus the initial stellar density (within the half-mass radius), $\rhoh$, that determines the efficiency of binary disruption in young star clusters: the larger the $\rhoh$, the more efficient binary burning gets ($\tcr$ decreases) and the lower the final binary fraction. The lowest final binary fraction achieved with their densest computed model ($\rhoh=7.55\times10^5\mpc$) after $5$~Myr of stimulated evolution is $\fb\approx30$~per~cent. In the period range observed by \citet{c05}, the binary fraction then still amounts to $\fb=23$~per~cent. If the retrograde population were to solely stem from the GC \ocen, its present-day density, $\rhoh\approx4.7\times10^3\mpc$ \citep[$M=4\times10^6\msun,\;\rh=6.44$~pc,][]{e1995wcen}, would thus not be sufficient to explain the virtual absence of binaries inside \ocen\  nor the deficiency of binaries on retrograde orbits (Sect.~\ref{sec:retrograde}). Instead, it had to have formed with a stellar density beyond the \citet{mko11} models. The difficulty here in matching the binary fraction of about $10$~per~cent on retrograde orbits is that the further the initial $\rhoh$ is increased, the stronger the initial cluster expansion gets through energy generation via binary burning \citep[Fig.~4 in][]{mko11}. This in turn rapidly reduces the binary burning efficiency to moderate levels. Therefore, extremely high stellar densities will be needed to explain the observations.
\subsection{\ocen\  as the host galaxy of a population of star clusters}
\label{sec:popmodel}
If the retrograde population were to solely stem from the star cluster population of \ocen's former host, the cumulative numbers of single stars and binaries from all star clusters would need to result in a low binary fraction. Due to dynamical effects such as the energy generated in binary encounters near the cluster core and residual-gas expulsion, star clusters expand and eventually dissolve to build up a galaxy's field population. This is similar to the idea put forward in \citet{mk11}, that the stellar single and binary star content of the Milky Way field population is the sum of stars and binaries that were born and processed in now dissolved star clusters of different initial density and, thus, with different binary disruption efficiencies. This assumed and now dissolved population of star clusters were the building blocks of \ocen\  in the present scenario.

A low binary fraction can only be achieved in this scenario if most systems on retrograde orbits originate from dense star clusters, which are efficient in destroying binaries. Neither among present-day nor young star clusters does there currently seem to be a (strong) correlation between mass and radius for young star clusters \citep{zepf1999,larsen2004,scheepmaker2007,mk12}. This implies that stellar density increases for young star clusters with increasing stellar mass, along with the binary destruction efficiency (Sect.~\ref{sec:singrequire}). Since typical star cluster populations initially follow a steep power-law ECMF \citep{ll2003,fuentemarcos2004,gieles2006},
\begin{equation}
 \xi_{\rm ECMF}\propto M_{ecl}^{-\beta}\;,\quad\beta\approx2\;,
 \label{eq:xiecmf}
\end{equation}
there are few massive clusters that efficiently burn binaries compared to those with a low mass and a low degree of binary dissolution. However, more systems ($\ncms$ in Eq.~\ref{eq:fb}) form in massive clusters than in low-mass clusters. \citet{mk11} demonstrate that for a Milky Way-type galaxy most binaries in the Galactic field originate from low-mass clusters ($<300\msun$, stimulated evolution not efficient), while most single stars are donated by massive clusters ($>10^4\msun$, stimulated evolution efficient), resulting in a global binary fraction (including all stellar masses over all periods) of $40-50$~per~cent, as observed in the field \citep{DuqMay91,r2010}. In the accessible range of periods in the spectroscopic observations by \citet{c05}, this corresponds to $\fb\approx21$~per~cent according to the \cite{mk12} model.

A means to reduce the binary fraction is to increase the star formation rate (SFR) at which the star cluster population formed. This is because the SFR determines the mass of the most massive cluster, $\meclmax$, that is able to form in a population of star clusters via the $\meclmax$--SFR relation \citep{w2004,k2013rev,randriamanakoto2013},
\begin{equation}
 \frac{\meclmax(\rm SFR)}{\msun}=84793\times\left(\frac{\rm SFR}{\myr}\right)^{0.75}\;.
 \label{eq:meclmax}
\end{equation}
This relation in fact follows from equating the total mass of stars formed in all embedded clusters (Eq.~\ref{eq:xiecmf}) over the time $\delta t \approx 10$~Myr to SFR$\cdot \delta t$ \citep{w2004}. Thus, the higher the SFR, the larger the mass of the most massive cluster. Due to the abovementioned lack of a strong mass-radius relation, producing more massive clusters is equivalent to producing denser clusters. And in a denser environment, binary processing is more efficient. Increasing the SFR thus results in more single stars being donated to the integrated population and reduces the binary fraction in it. In particular, \citet{mk11} show that a SFR of $10^3\myr$ ($\Rightarrow\meclmax=1.5\times10^7\msun$) reduces the global binary fraction to $30-40$~per~cent in elliptical galaxies. This corresponds to $\fb\approx18$~per~cent in the period range accessible by \citet{c05} and is thus still insufficient to explain the deficiency of binaries on retrograde orbits.

Nevertheless, four ways remain  to further lower the binary fraction: (i) increasing the SFR even further; (ii) making star clusters more compact to enhance binary burning in all clusters (via reducing $\rh$ for clusters); (iii) adding more massive (i.e. dense) star clusters, each contributing many single stars (via a flatter ECMF as a result of a lower index $\beta$ in Eq.~\ref{eq:xiecmf}); and (iv) suppressing the formation of low-mass clusters in the population, which would otherwise contribute too many binaries (via increasing the minimum cluster mass, $\meclmin$, of the population) -- or a combination of the four.

Suppression of low-mass cluster formation might occur in the case of a strong starburst with a high SFR. This was seen for the first time in self-consistent chemo-dynamical computations of the formation of tidal-dwarf galaxies \citep{p2014}. A local peak in the SFR history (an increase from $\approx10^{-4}\myr$ to $\approx1\myr$ over a period of $\approx50$~Myr in their models) produced very massive clusters ($M\gtrsim10^4\msun$) in the dense centre of the dwarf and produced about the same amount of low-mass clusters ($M\lesssim100\msun$) in the outskirts, with star cluster formation in the intermediate-mass regime being completely suppressed (Ploeckinger, private communication). Suppression of low-mass cluster formation (via increasing $\meclmin$) and a high SFR (resulting in larger $\meclmax$) might thus go hand in hand.

\subsection{Gauging the model against the prograde binary population}
\label{sec:gauge}
The integrated binary population models in \citet{mk11} have been proven to successfully reproduce the properties of the Galactic field M- and G-dwarf binary sub-populations as well as for the whole field population integrated over all spectral types, simultaneously matching their observed binary fractions, orbital period, mass ratio, and eccentricity distributions with a single set of parameters (SFR=$3\myr$, $\meclmin=5\msun$, $\beta=2$, $\rh=0.1-0.2$~pc). This allowed the prediction of binary populations in elliptical and dwarf irregular galaxies.

\citet{c05}'s estimated binary fraction for their prograde sub-population is consistent with the Galactic field data acquired by \citet{DuqMay91} within the mutual error bars ($\fb=26\pm2$ vs. $\fb=21\pm3.5$~per~cent). The \citet{c05} binary fraction is, however, closer to the suggested unevolved universal initial period distribution \citep[Fig.~\ref{fig:pdists}]{k95c,k95b}, which has about $25$~per~cent of all initial binaries in the indicated period range. While this might suggest that the orbital parameters of the population of spectroscopically selected binaries on prograde orbits are still primordial\footnote{This is not entirely unrealistic as binaries from this separation range are rather hard and more difficult to affect through dynamics than wider binaries (see Sect.~\ref{sec:hardsoft}).}, the fraction of binaries should nevertheless be lower than the initial value in this range, even if these were not affected by dynamics at all. This comes about as the production of single stars through the disruption of longer-period (soft) binaries dilutes $\fb$ for harder binaries as well (as discussed in Sect.~3.2 and Marks et al. 2014).

While being consistent, a higher-than-initial binary fraction in the accessible period range, if it were real, would not easily be reconciled with the suggested universal IBP. While new hard binaries can in principle form dynamically in the dense cores of GCs \citep{fir2009}, they probably do not contribute much to a cluster's population \citep{hut1992}. Furthermore, as the suggested IBP consists of $100$~per~cent binaries, the soft ones need to be broken up (lowering the hard binary fraction as well) before new hard binaries can form in any case. In what follows, however, the universal IBP \citep{k95c,k95b} will be used.

\section{Results}
\subsection{\ocen\  as a single cluster}
\label{sec:ocenclust}
The initial stellar density required to lower the binary fraction for systems on retrograde orbits about the Milky Way to $\approx10$~per~cent in the relevant period range is found to be $\rhoh\approx5\times10^8\mpc$ using the \citet{mko11} models. With \ocen's present-day mass ($M=4\times10^6\msun$), its half-mass radius had to have been $\approx0.1$~pc. This is significantly smaller than its present-day half-mass radius of about $6.44$~pc, but a strong initial phase of expansion is expected through binary burning \citep{mko11} and residual-gas expulsion \citep[e.g.][]{pfalzner2014,bk2017}. Its small initial size would be in line with the observed compact formation of young star clusters \citep{mk12}. Allowing additionally for early significant mass loss over its tidal boundary through cluster expansion, \ocen\  might have been much more massive than observed nowadays. In order for a cluster with initial mass\footnote{For stellar populations, the stellar density peaks at a mass of $\approx10^8\msun$, beyond which the two-body relaxation time becomes larger than a Hubble time and dark matter seems to appear \citep{dhk2008}. Therefore, $10^8\msun$ is used as the maximum possible mass that \ocen\  could have had if it were a single cluster.} $M=10^8\msun$ to have the same density ($\rhoh\approx5\times10^8\mpc$), that is, the same degree of binary disruption efficiency, the initial cluster size might have been as large as $\rh\approx0.3$~pc. In both cases, for an (unrealistic) homogeneous density distribution, the average binary-to-binary distance would have been $\approx380$~AU between their centres of mass.

\subsection{\ocen\  as the host galaxy of a population of star clusters}
\label{sec:ocendwarf}
\begin{figure}
 \includegraphics[width=0.48\textwidth]{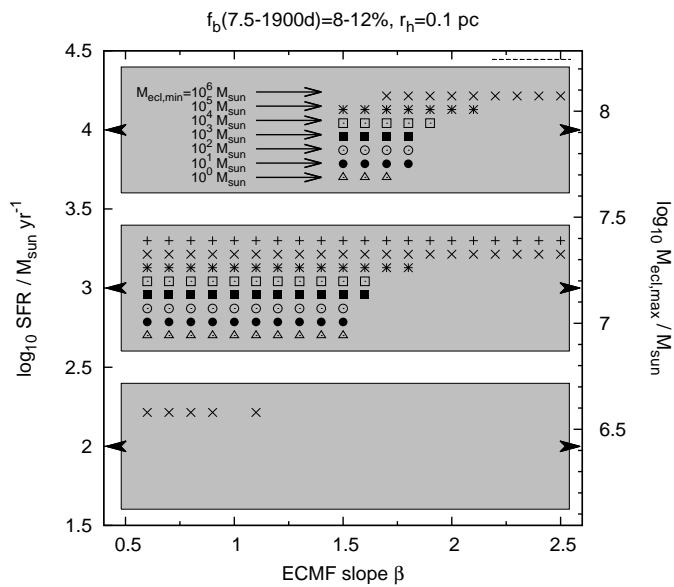}
 \caption[Solutions to the binary-deficient retrograde population]{Each symbol in the SFR vs. ECMF slope $\beta$ (Eq.~\ref{eq:xiecmf}) plane corresponds to a solution that reproduces the low binary fraction for the retrograde stellar population if typical star cluster half-mass radii were only $\rh=0.1$~pc. All symbols inside one of the grey boxes have the SFR indicated by the arrow on the left of each box. Each row of symbols in a box corresponds to the minimum mass of the cluster ($\meclmin$) that was able to form. The SFRs on the left ordinate correspond to a maximum cluster mass, $\meclmax$, on the right ordinate using Eq.~(\ref{eq:meclmax}). The binary fraction for the retrograde population is $\approx$10~per~cent if the SFR in the \ocen\  galaxy was large, the ECMF has been top-heavy (low~$\beta$), and low-mass star cluster formation was suppressed (large~$\meclmin$).}
 \label{fig:paramspace}
\end{figure}
In order to find solutions for which the binary fraction in the relevant period range reduces to $\approx10$~per~cent, a grid over the model parameters ($\rh$, $\beta$, $\meclmin$, SFR) was run using \bipos~\citep*{dmk2021}. Figure~\ref{fig:paramspace} shows that solutions can only be obtained if the SFR in \ocen\  has been large (SFR$\gtrsim10^3\myr$). Furthermore, only in a population of very compact star clusters (typical $\rh\approx0.1$~pc) can the binary fraction be lowered to the observed degree. Such small cluster half-mass radii are consistent with the typical initial size for star clusters that contributed to the Milky Way's (prograde) field population as derived in \citet{mk11}. Allowing for larger cluster radii (i.e. less efficient destruction of binaries) hardly\footnote{There are a few solutions for $\rh=0.2$~pc, SFR$=10^4\myr$ and $\beta<1.5,$ but not for larger radii.} reduces the binary fraction between $P=1.9$~and~$7500$~days for any set of the parameters.
If the lowest-mass clusters in \ocen\  that contributed to the retrograde field population were comparable to the lowest-mass stellar groups visible today in the Milky Way \citep[$\meclmin\approx5\msun$ for Taurus-Auriga type of group;][]{jd2018}, a top-heavy ECMF ($\beta\lesssim2$) would be required in addition to high SFRs and compact star clusters (filled circles and open triangles in Fig.~\ref{fig:paramspace}). This becomes necessary as an increased number of high-mass (i.e. high-density) clusters is needed in order to feed the field population with single stars to compensate for the many binaries retained in low-mass clusters (Sect.~\ref{sec:popmodel}). For a SFR$=10^4\myr$, the binary fraction even decreases below $10$~per~cent if the ECMF had been too top-heavy ($\beta<1.6$) and would therefore not be a solution to the problem.

If in the \ocen\  galaxy the ECMF had been as observed in present-day young star cluster populations ($\beta\approx2$), low-mass cluster formation would be required to be significantly suppressed ($\meclmin\gtrsim10^6\msun$; crosses in Fig.~\ref{fig:paramspace}). If it were not, lower-mass clusters would contribute too many binaries to the field population, boosting the binary fraction above the observed value within $P=7.9-7500$~days.

\subsection{Prediction of properties for the Milky Way's retrograde binary population}
\begin{figure*}
 \includegraphics[width=0.45\textwidth]{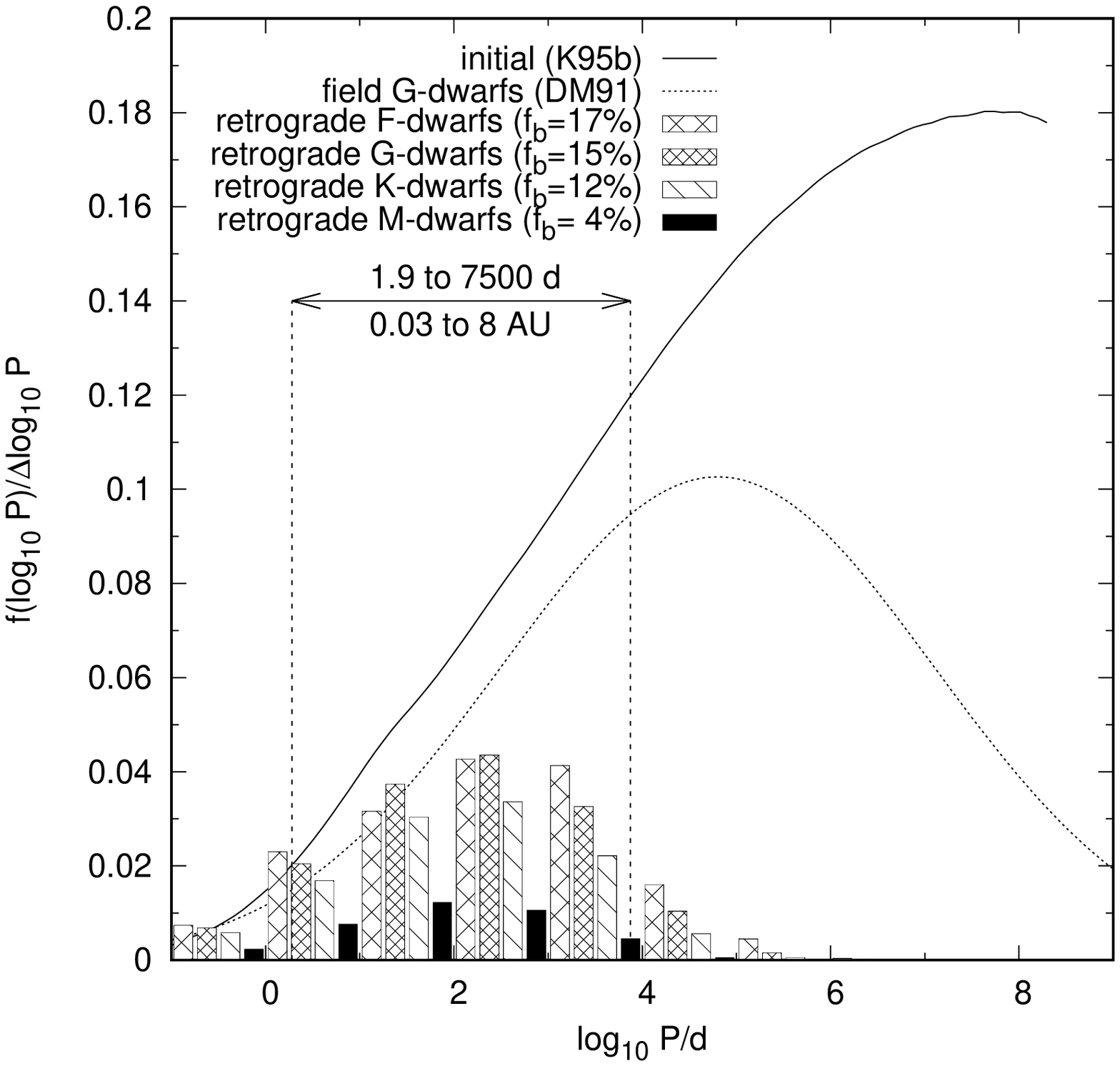}
 \includegraphics[width=0.45\textwidth]{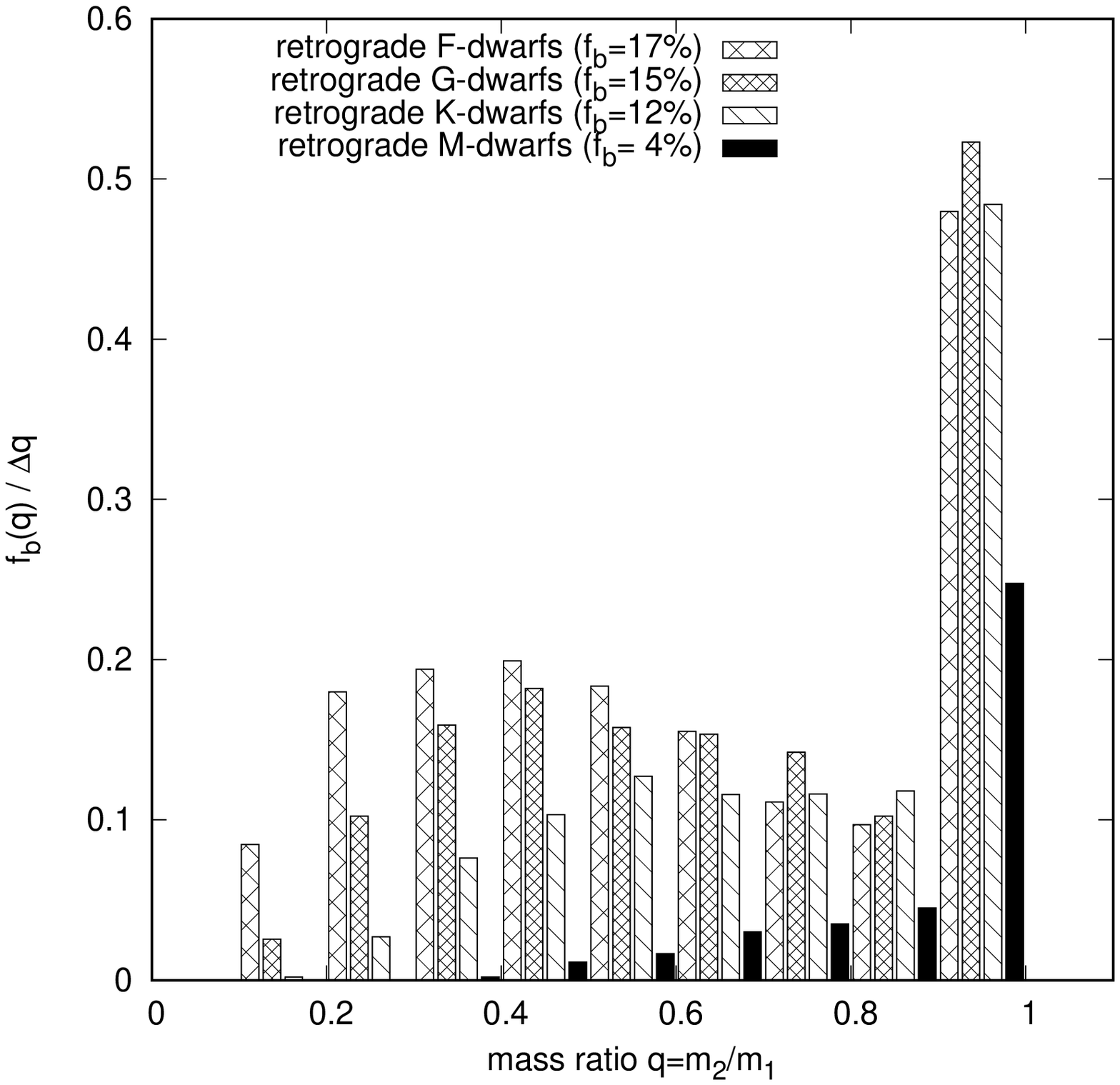}
 \caption[Prediction of period and mass-ratio distributions for retrograde binaries]{Prediction of the binary period distribution (left panel) and mass-ratio distribution (right panel) for field F, G, K and M dwarfs on retrograde orbits. Expected binary fractions over the whole period range depend on spectral type and are indicated in the key. Every four neighbouring histogram bars with different filling patterns are in the same bin, whose centre is located in the middle between two integers in $log_{10} P/d$. For example, the centre of each of the four bars between $log_{10} P/d=2$ and $3$ is located at $log_{10} P/d=2.5$ but is shifted against the others for the sake of visibility. While all sub-populations share the same initial period distribution (left panel, solid line), a common initial distribution for the mass ratio cannot be shown here since its shape depends on the spectral type \citep[e.g. rising for M dwarfs over the whole q range but declining for F and G dwarfs for $q \gtrapprox 0.4$;][]{mk11}. Thus, for example, $\approx1\%$ of all M dwarfs in the retrograde population are binaries with $\log_{10}P/d=2-3$ (left panel, black filled bar). All the retrograde M-dwarf binaries have a mass-ratio distribution shown by the black histogram in the right panel.}
 \label{fig:predictions}
\end{figure*}
Identifying possible solutions to the deficiency of binaries on retrograde orbits using the \citet{mko11} and \citet{mk11} models within the \bipos~code \citep*{dmk2021} allows the prediction of the properties that a complete sample, in which all binaries and single stars on retrograde orbits are identified, should match if \ocen\  is the origin of the retrograde stellar population and the models are applicable. Figure~\ref{fig:predictions} shows the then expected period and mass-ratio distribution for field F, G, K, and M dwarfs on retrograde orbits if the whole range of periods and masses were accessible to an observer. The distributions for the different possible parameter combinations identified in the previous section are statistically indistinguishable since the binary fraction fully determines the shape of the distribution in the model.

The period distribution retains a bell-shaped form as observed for the prograde field population, although with a significantly lower binary fraction. The binary fraction depends on spectral type and is indicated in Fig.~\ref{fig:predictions} \citep[$f_b(F)>f_b(G)>f_b(K)>f_b(M)$;][]{mk11}.
The mass-ratio distribution is rising for field M dwarfs over the full q range but is flat for K dwarfs and declining for G and F dwarfs if $q \gtrapprox 0.4$. A peak at high mass ratios is seen due to eigenevolution within proto-binary discs, which tends to equalise the component masses due to mass transfer at aphelion. The different shapes of the mass-ratio distribution stemming from the universal IBP depending on spectral type have been discussed and explained in \citet{mk11} for the prograde field population, and they do indeed match observations.

\section{Conclusion and discussion}
It has been suggested that \ocen\  is the source of the observed binary deficiency for systems on retrograde orbits \citep{c05}. Possible star formation conditions have been identified that would lead to the observed low binary fraction. The present analysis made use of the publicly available \bipos~\citep*{dmk2021}, which starts with a universal IBP \citep{k95c,k95b} and calculates the resulting binary properties after dynamical processing inside a single cluster or in a population of star clusters that dissolved to become part of the field (see the appendix for its usage in this paper). The code is based on the works by \citet{mko11} and \citet{mk11}.

In the two scenarios explored, \ocen\  is either considered to be (i) the most massive single star cluster in the Milky Way or (ii) the surviving core of a former dwarf galaxy that was host to a population of star clusters, each of which evolves its own binary population individually before dissolving and building the low-metallicity retrograde field population associated with \ocen.

In the former scenario, \ocen\  is required to have been very dense, its birth stellar peak density being of the order of $\rhoh\approx10^8\mpc$. Only with such a high density was the binary destruction efficiency sufficiently large to lower the binary fraction to the observed value, $\fb=10$~per~cent, in the period range $P=1.9-7500$~days. With \ocen's present-day mass, its initial half-mass radius would need to have been as small as $0.1$~pc, consistent with observed sub-parsec sizes of young star clusters. If its progenitor had been more massive, an initial half-mass radius of up to $\rh=0.3-0.4$~pc would have been possible.

In the latter scenario, the star cluster population in the \ocen\  dwarf galaxy had formed in an intense starburst, with SFRs exceeding $1000\myr$. Star formation rates up to $\approx3200\myr$ have been observed for some far-IR, bright, high-redshift quasars in the sample of \citet{nguyen2020}. Some solutions that reproduce the retrograde binary population additionally suggest that its ECMF has been top-heavy ($\beta\lesssim2$ in Eq.~\ref{eq:xiecmf}) and that low-mass star cluster formation was suppressed, that is, the mass of the lowest-mass clusters in the population had been significantly larger than the lowest-mass group present today in the Milky Way ($\approx5\msun$ for a Taurus-Auriga type of group). This ensured that massive, dense clusters, in which binary disruption is efficient, were the dominant donators of systems to the field.

We caution that very high densities in the GC progenitor scenario (Sect.~\ref{sec:ocenclust}) as well as very high SFRs in the dwarf galaxy progenitor scenario (Sect.~\ref{sec:ocendwarf}) might alter star formation conditions significantly and lead to top-heavy IMFs in star-forming regions \citep{papa2010,mkdp12}, which are not included in \bipos.

Binary period and mass-ratio distributions for late-type stars on retrograde orbits are predicted according to the here-constrained star formation conditions for \ocen. The period distributions are bell-shaped, as are the ones in the field \citep{DuqMay91,r2010}. Overall, binary fractions (i.e. integrated over the full range of periods and mass ratios) range from $4$~per~cent for M dwarfs to $18$~per~cent for F dwarfs. This then results in the observed value of $10$~per~cent for binaries on retrograde orbits in the surveyed mass and period range. The shape of the mass ratio distribution depends on spectral type, as explained before by \citet{mk11}. Not considering a peak at high mass ratios due to eigenevolution (see the appendix), the mass-ratio distribution is rising for M dwarfs, rather flat for K dwarfs, and declining for F and G dwarfs for $q>0.4$.

Given this study and the cumulating evidence in the literature, the dwarf galaxy scenario for \ocen\  is tentatively preferred here. Validating the period and mass-ratio distributions predicted here for retrograde binaries about the Milky Way in future observations would both support the notion that \ocen\  and the retrograde binary population are indeed connected and increase confidence in the binary population synthesis model that is at the core of \bipos.

\begin{acknowledgements}
The authors thank Bruce W. Carney and David Latham for providing the data-files used for the analysis in \citet{c05} as well as help with their data. Pavel Kroupa and Jörg Dabringhausen acknowledge support from the Grant Agency of the Czech Republic under grant number 20- 21855S.
\end{acknowledgements}

\appendix
\section{Usage of the Binary Population Synthesiser}
\subsection{General remarks}
The \bipos~is capable of (i) calculating  an evolved binary population, after the \cite{k95c,k95b} IBP with 100~per~cent binaries at birth, which adjusted their orbital parameters in the gas disc in which they formed (eigenevolution), have gone through up to 5~Myr of stimulated evolution in a young star cluster using the analytical method described in \citet{mko11}, or (ii) producing a galaxy-wide field-binary population using the math outlined in \citet{mk11}. Results are available within seconds.

\bipos~calculates binary fractions and outputs binary OPDs for periods, semi-major axes, apparent separations on the sky, mass ratios, eccentricities, angular momenta, binding energies, and orbital time and phases, ready for comparison with observed OPDs and the prediction of properties in unknown binary populations. Furthermore, it may be used to generate `birth', `initial' (after eigenevolution), and `dynamically evolved' binary populations for use in numerical computations, thus avoiding the time-consuming initial phase of a simulation with large IBPs.

The code and its usage is described in \citet*{dmk2021} and is downloadable from GitHub\footnote{Download from \texttt{https://github.com/JDabringhausen/BiPoS1}}. For a quick start, you can compile the program using {\verb make } and call
\begin{lstlisting}
./BiPoS
\end{lstlisting}
for the help menu. Information regarding the application of \bipos~in this paper is given in Appendix~\ref{sec:biposusage}.

\subsection{Generating binary fractions and OPDs in this paper}
\label{sec:biposusage}
\bipos~requires a library of binaries and their orbital parameters to be generated as a basis to work with. To write a statistical sample of $10^6$ binaries into the file \verb retrograde.dat, you type
\begin{lstlisting}
./BiPoS genlib Nlib=1000000
        libname=retrograde.dat
\end{lstlisting}
in the command line, which outputs a library of binaries with periods selected from the \cite{k95c,k95b} IBP with masses ranging between $0.08\msun$ and $150\msun$ from the \cite{k2001} IMF. Although \bipos~would allow it, the library itself is not restricted here to the mass range of stars observed by \cite{c05} since it is meant to represent the composition of the \ocen\  population at birth. The more massive progenitor of \ocen\  likely contained stars from the full spectrum of stellar masses. The restriction to the observed binaries is done in a later step.

The number of binaries in the library, in contrast, does not need to represent the number of binaries at birth. It is of statistical importance only. For common use cases: the larger the number of binaries in this library, the better. A larger number will increase the computation time in the following steps because the library is read by \bipos~every time a calculation is performed. A lower number will increase statistical differences between the results produced by \bipos~for different realisations of this library.
In Sect.~\ref{sec:popmodel} the Milky Way's Galactic field-binary fraction for prograde binaries is calculated in the period range $\lP (days)=0.3-3.9$ and masses $0.4-1.0\msun$ as targeted in the Carney-Latham survey \citep{c05} using the galactic field synthesis mode ({\verb ./BiPoS } {\verb field } {\verb help }) in \bipos. With a SFR of $3\myr$, an ECMF slope $\beta=2$, a minimum cluster mass $\meclmin=5\msun$, and a typical embedded cluster half-mass radius of $\rh=0.1$~pc for the Milky Way \citep{mk11}, the command
\begin{lstlisting}
./BiPoS field rh=0.1 SpT=u mmin=0.4
        mmax=1.0 constrain=P,0.3,3.9,x
        libname=retrograde.dat
\end{lstlisting}
outputs the stated binary fraction of about 21~per~cent in the terminal. An arbitrary number can be entered for {\verb x } in the {\verb constrain } parameter, or it may be left as is since the code will ignore it in this use case.\footnote{The value for x determines the number of equal-sized bins there would be if the period OPD were actually produced. To only determine the binary fraction, the binning is irrelevant.} In order for a user-defined mass range to be set, one needs to add {\verb SpT=u } alongside the parameters {\verb mmin=... } and {\verb mmax=... }, where the masses have to be given in $\msun$. We note that the \bipos~parameters for the ECMF slope (\verb beta ), the SFR (\verb sfr ), and the minimum embedded cluster mass (\verb meclmin ) do not need to be set in this case since these Milky Way values  are the standard values in \bipos~\citep[cf.][]{mk11}. To calculate the corresponding number for an elliptical galaxy with a SFR of $1000\myr$, you simply add {\verb sfr=1000 } to the above command and arrive at about 18~per~cent, the number given in this paper.

In Sect.~\ref{sec:gauge} the fraction of binaries in the IBP in the \cite{c05} mass and period range is calculated to be 25~per~cent by adding the {\verb +init } command-line parameter:
\begin{lstlisting}
./BiPoS field +init rh=0.1 SpT=u mmin=0.4
        mmax=1.0 constrain=P,0.3,3.9,x
        libname=retrograde.dat
\end{lstlisting}

As stated in Sect.~\ref{sec:ocenclust}, a 10~per~cent binary fraction as required by the observations is approximately arrived at for a density of $\rhoh\approx5\times10^8\mpc$ in an individual GC (e.g. by using the present-day mass $M=4 \times 10^6\msun$ for \ocen\  and a former half-mass radius of $\approx0.1$~pc). Using the \bipos~cluster binary population synthesis mode ({\verb ./BiPoS } {\verb clust } {\verb help }), the full command reads
\begin{lstlisting}
./BiPoS clust mecl=4000000 rh=0.1
        SpT=u mmin=0.4 mmax=1.0
        constrain=P,0.3,3.9,x
        libname=retrograde.dat
\end{lstlisting}

Assuming a much more massive progenitor of $\mecl=10^8\msun$, one can allow for a three-times larger half-mass radius of $\approx0.3$~pc:
\begin{lstlisting}
./BiPoS clust mecl=100000000 rh=0.3
        SpT=u mmin=0.4 mmax=1.0
        constrain=P,0.3,3.9,x
        libname=retrograde.dat
\end{lstlisting}

In order to find parameter sets for the Galaxy field synthesis mode in \bipos~that lead to the observed binary fraction of $\approx10$~per~cent (Sect.~\ref{sec:ocendwarf}), an adapted version was used to scan the large parameter space. This approach is in principle equivalent to subsequent calls of the form
\begin{lstlisting}
./BiPoS field beta=1.9 sfr=10000
        meclmin=100000 rh=0.1
        SpT=u mmin=0.4 mmax=1.0
        constrain=P,0.3,3.9,x
        libname=retrograde.dat
\end{lstlisting}
and varying parameters {\verb beta }, {\verb sfr }, and {\verb meclmin }.

To output the raw data for the histograms in Fig.~\ref{fig:predictions} that predict the period- and mass-ratio OPD for retrograde binaries in the Galactic field, one can either use the cluster mode or the field mode in \bipos, depending on whether the GC or dwarf galaxy progenitor scenario is looked at. As long as the correct values for the parameters that lead to the observed binary fraction in the restricted mass and orbital period range accessible by \cite{c05} are used, the resulting binary fraction alone will determine the distributions uniquely. Here, the field synthesis mode is used. The command
\begin{lstlisting}
./BiPoS field beta=1.9 sfr=10000
        meclmin=100000 rh=0.1 OPD=Pq
        SpT=FGKM libname=retrograde.dat
\end{lstlisting}
outputs eight individual files with the period (P) and mass ratio (q) OPD for stars of spectral type F, G, K, and M. The \cite{c05} period range no longer enters here via the {\verb constrain=... } command-line parameter, since for the prediction in Fig.~\ref{fig:predictions} the whole range of periods is of interest.

We note that the OPDs might look slightly different if other solutions for the parameters {\verb beta }, {\verb sfr }, and {\verb meclmin } are used in the field synthesis mode or if \bipos~is run in the cluster population synthesis mode. The OPDs are, however, statistically indistinguishable.

In order to see the standard mass ranges adopted by \bipos~for each spectral type, you call
\begin{lstlisting}
./BiPoS SpT help
\end{lstlisting}
The option to choose an individual mass range by the user, setting the {\verb SpT=u } parameter together with {\verb mmin=... } and {\verb mmax=... } as above, remains unaffected by these standard mass ranges.

\bibliographystyle{aa}
\bibliography{biblio}
\makeatletter   \renewcommand{\@biblabel}[1]{[#1]}   \makeatother

\label{lastpage}

\end{document}